\documentclass[referee,usenatbib]{mn2e} 
\usepackage{epsfig}

\newcommand{\beq}{\begin{equation}} 
\newcommand{\eeq}{\end{equation}}

\newcommand{\omp}{\mbox{$\Omega_p$}} 
\newcommand{\bfr}{\mbox{\boldmath $r$}} 
 
\newcommand{\bfv}{\mbox{\boldmath $v$}}
\newcommand{\del}{\mbox{\boldmath $\nabla$}}
\newcommand{\cross}{\mbox{\boldmath $\times$}}
\newcommand{\p}{\mbox{$\partial$}}
 
\newcommand{\kmspc}{{\rm\,km\,s^{-1}\,pc^{-1}}}

\def\etal{et al.}
\newcommand{\aanda}{\rm A\&A} 
\newcommand{\mnras}{\rm MNRAS}
\newcommand{\apj}{\rm ApJ}
\newcommand{\aj}{\rm AJ}

\title[Recovery of transverse velocities]
{Recovery of transverse velocities of steadily rotating patterns in flat
galaxies}

\author[Sridhar \& Sambhus]
{S. Sridhar$^1$\thanks{ssridhar@rri.res.in} and Niranjan Sambhus$^2$
\thanks{sambhus@astro.unibas.ch} \\
$^1$Raman Research Institute, Sadashivanagar, Bangalore 560080, India \\
$^2$Astronomisches Institut, Universit\"at Basel, Venusstrasse 7, 
CH--4102 Binningen, Switzerland}

\begin{document}

\maketitle

\begin{abstract}
The transverse velocities of steadily rotating, non--axisymmetric patterns
in flat galaxies may be determined by a purely kinematical method, using
two dimensional maps of a tracer surface brightness and radial current
density. The data--maps could be viewed as the zeroth and first
velocity moments of the line--of--sight velocity distribution, which is 
the natural output of integral--field spectrographs. Our method is closely
related to the Tremaine--Weinberg method of estimating pattern speeds of
steadily rotating patterns, when the tracer surface brightness satisfies a
source--free continuity equation. We prove that, under identical
assumptions about the pattern, two dimensional maps may be used to recover
not just one number (the pattern speed), but the full vector field of
tracer flow in the disc plane. We illustrate the recovery process by
applying it to simulated data, and test its robustness by including the
effects of noise. 
\end{abstract} 

\begin{keywords}
galaxies: kinematics and  dynamics---galaxies: nuclei
\end{keywords}

\section{INTRODUCTION}

Over the past decade long--slit spectrographs have given way to
integral--field spectrographs (IFS), which produce spectra over a fully
sampled, two dimensional region of the sky \citep[see e.g.][]{bac01, tha01,
eb02}. These spectral maps (also called the line--of--sight velocity
distribution, hereafter LOSVD) contain important information on the
flow patterns of non--axisymmetric features in galaxies and their nuclei.
It is widely believed that bars and spiral patterns in disc galaxies
could influence galaxy evolution through their roles in the transport of
mass and angular momentum. These processes are not understood completely,  
and IFS maps  might be expected to play a key role in the construction of
dynamical models of evolving galaxies \citep{dz02, e02}. A limitation is
that IFS maps provide information about radial, but not transverse,
velocities. It is not possible to recover the unmeasured transverse
velocities without additional assumptions; a classic example is the
modelling of the warped disc of M83, using tilted, circular rings
\citep{rlw74}. However, the flows in non--axisymmetric features, such as
bars, are expected to be highly non circular, and a different approach is
needed. 

\citet[hereafter TW84]{tw84} considered steadily rotating patterns in 
flat galaxies, and showed how data from long--slit spectrographs may be 
used to estimate the pattern speed. Their method assumes that the disc of 
the galaxy is flat, has a well--defined pattern speed, and that the tracer
component obeys a source--free continuity equation. The goal of this paper
is to prove that, making identical assumptions about the pattern, IFS data
can be used to determine not just one number (the pattern speed), but the
transverse velocities, and hence the entire two dimensional vector field
of the tracer flow. Like the TW method, one of the strengths of our method
is that it is kinematic, and not based on any particular dynamical
model. Our main result is equation~(\ref{xcurrentobs}) of \S~2, which
provides an explicit expression for the transverse component of the tracer
current in the disc plane, in terms of its surface brightness and the
radial current density maps on the sky. This formula is applied in \S~3 to
a model of the lopsided disc in the nucleus of the Andromeda galaxy (M31), 
where we also discuss the effects of noise on the data--maps. \S~4 offers
conclusions.

\section{THE RECOVERY METHOD}

An IFS dataset consists of a two dimensional map of the luminosity
weighted distribution of radial velocities, the LOSVD. The LOSVD can
be regarded as a function of the three variables, $(X, Y, U)$, where
$X$ and $Y$ are cartesian coordinates on the sky, and $U$ is the radial 
velocity. The zeroth moment of the LOSVD over $U$ is $\Sigma_{\rm
sky}(X, Y)$, the surface brightness distribution on the sky, and the first
moment is $F_{\rm sky}(X, Y)$, the radial current density on the
sky\footnote{The mean radial velocity is then given by $\overline{U}(X,
Y)\,=\, (F_{\rm sky}/\Sigma_{\rm sky})\,$: the contour map of
$\overline{U}(X,Y)$ is often refered to as a ``spider diagram''.}.
Following TW84, we consider a thin disc that is confined to the $z\,=\,0$
plane, with $x$ and $y$ being cartesian coordinates in the disc plane. The
disc is inclined at angle $i$ to the sky plane ($i=0\degr$ is face--on,
and $i=90\degr$ is edge--on), with line of nodes coincident with the
$x$--axis. It is clear that the sky coordinates, $(X, Y)$, may be oriented
such that the $X$--axis and $x$--axis are coincident. Then $(X, Y) \,=\,
(x, y\cos i)\,$. 

The non--axisymmetric pattern of the tracer is assumed to rotate steadily
at angular rate, $\omp\,\hat{z}\,$. In this frame the continuity equation
for the tracer brightness assumes its simplest form. Let $\bfr$ be the
position vector in the rotating frame, $\Sigma(\bfr)$ the tracer surface
brightness, and $\bfv(\bfr)$ the streaming velocity field in the
inertial frame. An observer in the rotating frame sees the tracer move
with velocity, $\left[\bfv(\bfr) \,-\, \omp(\hat{{\bf z}} \cross
\bfr)\right]\,$. If the tracer brightness is conserved, $\Sigma$ and
$\Sigma\bfv$ must obey the continuity equation, $\del\cdot\left[
\Sigma\left(\bfv \,-\, \omp\hat{{\bf z}}\cross \bfr\right)\right] \;=\;
0\,$. Cartesian coordinates in the rotating frame may be chosen such that
they coincide instantaneously with the $(x, y)$ axis; thus $\bfr \,=\, (x,
y)\,$ and $\bfv(\bfr) \,=\, (v_x, v_y)\,$. In component form, the
continuity equation reads, 
\beq
\frac{\p\,(\Sigma v_x)}{\p x} \;+\; \frac{\p\,(\Sigma v_y)}{\p y} \;=\; 
\omp\left(x\frac{\p\Sigma}{\p y} \,-\, y\frac{\p\Sigma}{\p
x}\right)\,. 
\label{cont} 
\eeq 
\noindent which is equivalent to equations~(2) and (3) of TW84. The
quantities, $\Sigma(x, y)$ and $\Sigma(x, y)v_y(x, y)$, can be related
directly to the observed surface brightness and radial current density
maps: 
\begin{eqnarray} 
\Sigma(x, y) &\;=\;& \cos i\;\Sigma_{\rm sky}(X,
Y)\,,\\[1em] 
\Sigma(x, y)v_y(x, y) &\;=\;& \cot i\;F_{\rm sky}(X, Y)\,. 
\label{data} 
\end{eqnarray} 
\noindent Henceforth $\Sigma(x, y)$ and $\Sigma(x, y)v_y(x, y)$ will be
considered as known quantities. The unknowns in equation~(\ref{cont})
are $\omp$ and $\Sigma(x, y)v_x(x, y)$. Below we prove that both quantities
may be obtained by integrating over $x\,$.  We will assume that
$\Sigma(x, y)\,$, $\Sigma(x, y)v_y(x, y)$, and (the unknown quantity)
$\Sigma(x, y)v_x(x, y)$, all decrease sufficiently rapidly with distance,
such that all the integrals encountered below are finite.

Integrating equation~(\ref{cont}) over $x$ from $-\infty$ to
$x$, we obtain,
\beq
\Sigma(x, y)v_x(x, y) \;=\; -\frac{\p}{\p y}\int_{-\infty}^{x}dx'\,
\left(\,\Sigma v_y \,-\, \omp x'\Sigma\,\right)_{(x', y)} \;-\; \omp
y\Sigma(x, y)\,,
\label{xcurrent} 
\eeq 
\noindent where we have used $\Sigma(-\infty, y) = 0\,$, and 
$\Sigma(-\infty, y)v_x(-\infty, y) = 0\,$. We must also require that
$\Sigma(+\infty, y) = 0\,$, and $\Sigma(+\infty, y)v_x(+\infty, y) = 0\,$.
This leads to the condition,
\beq
\frac{\p}{\p y}\int_{-\infty}^{+\infty}dx\,
\left(\Sigma v_y \,-\, \omp x\Sigma\right) \;=\; 0\,.
\label{require} 
\eeq 
\noindent Since the integral in equation~(\ref{require}) is independent of
$y$, we can infer its value at large values of $|y|$. Therefore the
integral itself must vanish. i.e.
\beq
\omp\,\int_{-\infty}^{+\infty} dx\, x\Sigma(x, y) \;=\; 
\int_{-\infty}^{+\infty} dx\, \Sigma(x, y)\,v_y(x, y)\,,
\label{twrel} 
\eeq 
\noindent for {\em any} value of $y$. This will be recognised as the key
relation that TW84 employ to determine the pattern speed (see eqn.~5 of
their paper). As is clear from our derivation, the real significance of
equation~(\ref{twrel}) is an {\em eigenvalue} of equation~(\ref{xcurrent}). 
In other words, it provides a consistency condition that $\Sigma(x,
y)$ and $\Sigma(x,y)v_y(x, y)$ must satisfy, if $\Sigma(x,y)v_x(x, y)$ is
to be given by equation~(\ref{xcurrent}). Using equation~(\ref{data}), we
can rewrite equations~(\ref{twrel}) and (\ref{xcurrent}), such that $\omp$
and $\Sigma(x, y)v_x(x, y)$ are expressed directly in terms of observed
quantities\footnote{In principle, the value of $\omp$ given by
equation~(\ref{twobs}) should be independent of $Y$. However TW84
recommends multiplying equation~(\ref{twobs}) by some weight function,
$h(Y)\,$ and integrating over $Y$, to obtain an estimate of $\omp$ as the
ratio of two integrals---see equation~(7) of TW84.}:
\beq
\omp\sin i \int_{-\infty}^{+\infty}\,dX\,X\,\Sigma_{\rm
sky}(X, Y) \;=\; \int_{-\infty}^{+\infty}\,dX\,F_{\rm sky}(X, Y)
\,,\label{twobs}
\eeq
\beq
\left(\Sigma v_x\right)_{(x, y)} \;=\; -(\cos i)^2\frac{\p}{\p Y}
\int_{-\infty}^{X}dX'\,
\left(\frac{F_{\rm sky}}{\sin i} \,-\, \omp X'\Sigma_{\rm sky}\right)_{(X',
Y)} \;-\; \omp
Y\Sigma_{\rm sky}(X, Y)\,.
\label{xcurrentobs} 
\eeq
\noindent In the next section, equations~(\ref{twobs}) and
(\ref{xcurrentobs}) will be used on the simulated data of
Figure~(\ref{fig1}), to enable recovery of the entire two dimensional flow
vector field of a steadily rotating, lopsided pattern.

\section{APPLICATION TO SIMULATED DATA}

\begin{figure}
\centerline{\hspace*{0.2in}(a)\hspace{3.1in}(b)\hfill}
\centerline{\epsfig{file=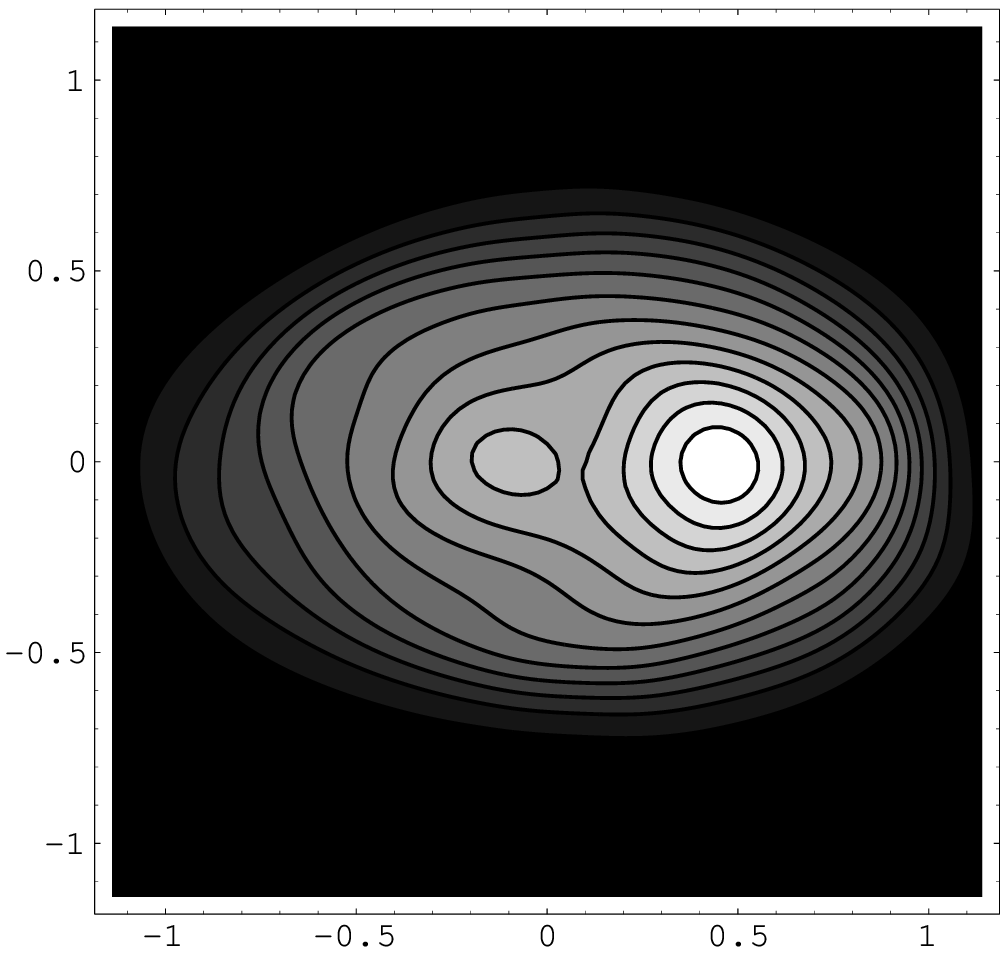,width=0.5\linewidth}
\epsfig{file=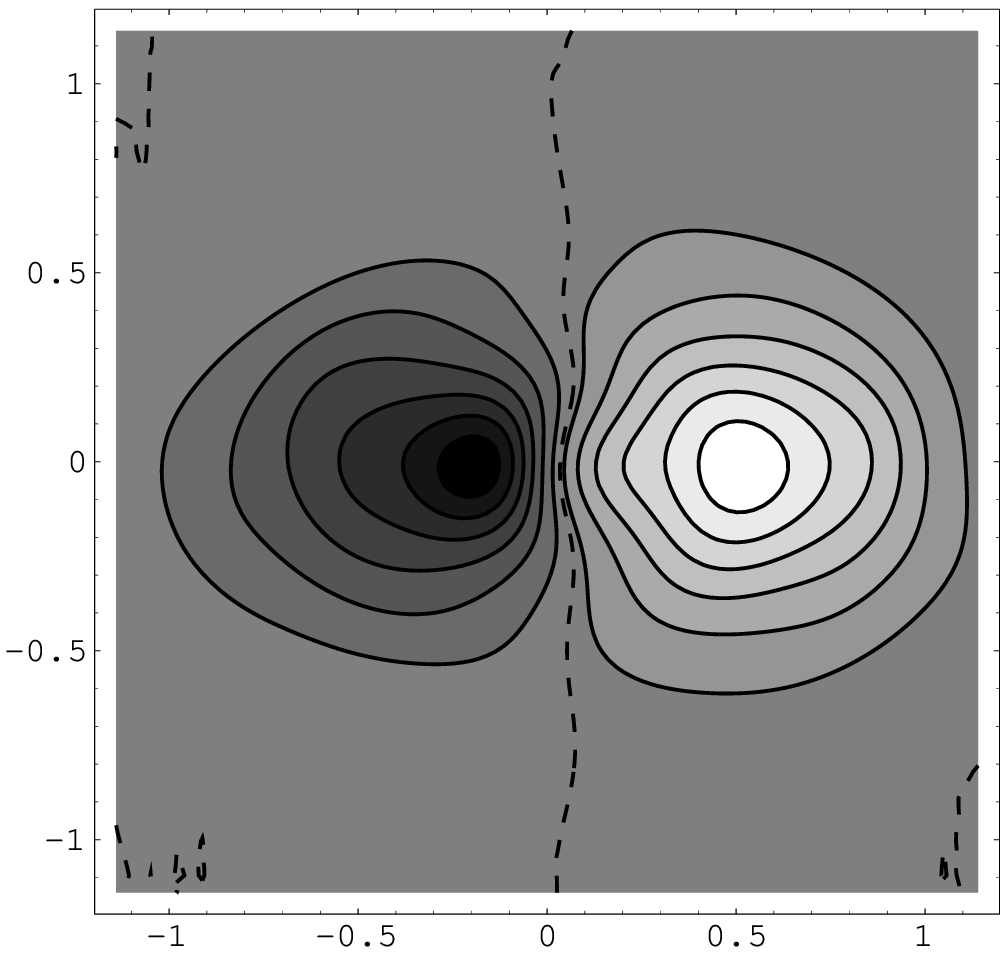,width=0.5\linewidth}}
\caption{Simulated data from a model disc inclined at $51\fdg54$: two
dimensional distribution of {\bf (a)} Surface brightness ($\Sigma_{\rm
sky}$) and {\bf (b)} Radial current density ($F_{\rm sky}$) of the model
disc.  The images have been smoothed with a circular Gaussian beam ($\sigma
= 0\farcs1$). The contour levels are arbitrary, but separated uniformly in
the values of $\Sigma_{\rm sky}$ and $F_{\rm sky}$, respectively.
In {\bf (b)} the black and white shadings correspond to negative
and positive radial current densities, and the dashed line is the zero radial
current density curve. In both maps, the line--of--nodes is along the $x$
axis. The axes scales are in ($''$). (``Data'' taken from SS02)}
\label{fig1}
\end{figure}

We used as data, simulated observations of a numerical model of the stellar
disc in the nucleus of the Andromeda galaxy (M31). A brief account of the
model is given below, and the reader is referred to \citet[hereafter
SS02]{ss02} for details. The nucleus of M31 is believed to harbour a Super
Massive Black Hole (SMBH) \citep{kb99}, surrounded by a dense stellar disc
which appears as a lopsided, double--peaked structure \citep{lau93}. The
two peaks are separated by about $0\farcs5$, with the fainter peak almost
coincident with the location of the SMBH \citep{lau98}. The dynamical
centre of the galaxy lies inbetween the two peaks, about $0\farcs1$ from
the SMBH. \citet{tre95} proposed that the SMBH was surrounded by an
eccentric disc of stars, whose orbital apoapsides were aligned in a manner
that gave rise to the lopsided peak in the density of stars. Our input
model is a dynamical model of this eccentric disc that was constructed by
SS02, based on the {\sl HST} photometry of \citet{lau98}. The model
consisted of about $230,000$ points distributed on a plane. Each point
(``star'') possessed five attributes: luminosity (or mass), location in the
plane, and two components of velocity. The lopsided pattern formed by
these points rotated steadily about an axis normal to the plane with a
(prograde) pattern speed equal to $16 \kmspc$; thus the model satisfied
all the assumptions used in \S~2. SS02 estimated an inclination angle $i =
51\fdg54$, and we use this value while projecting the model disc to the
sky-plane. To obtain a smooth distribution, we ``observed'' the model with
a circular Gaussian beam of $\sigma = 0\farcs1$~. Figure~(\ref{fig1}) shows
the surface brightness ($\Sigma_{\rm sky}$) and the radial current density
($F_{\rm sky} $). The line of nodes is along the $x$--axis.

\begin{figure}
\centerline{\hspace*{0.2in}(a)\hspace{3.1in}(b)\hfill}
\centerline{\epsfig{file=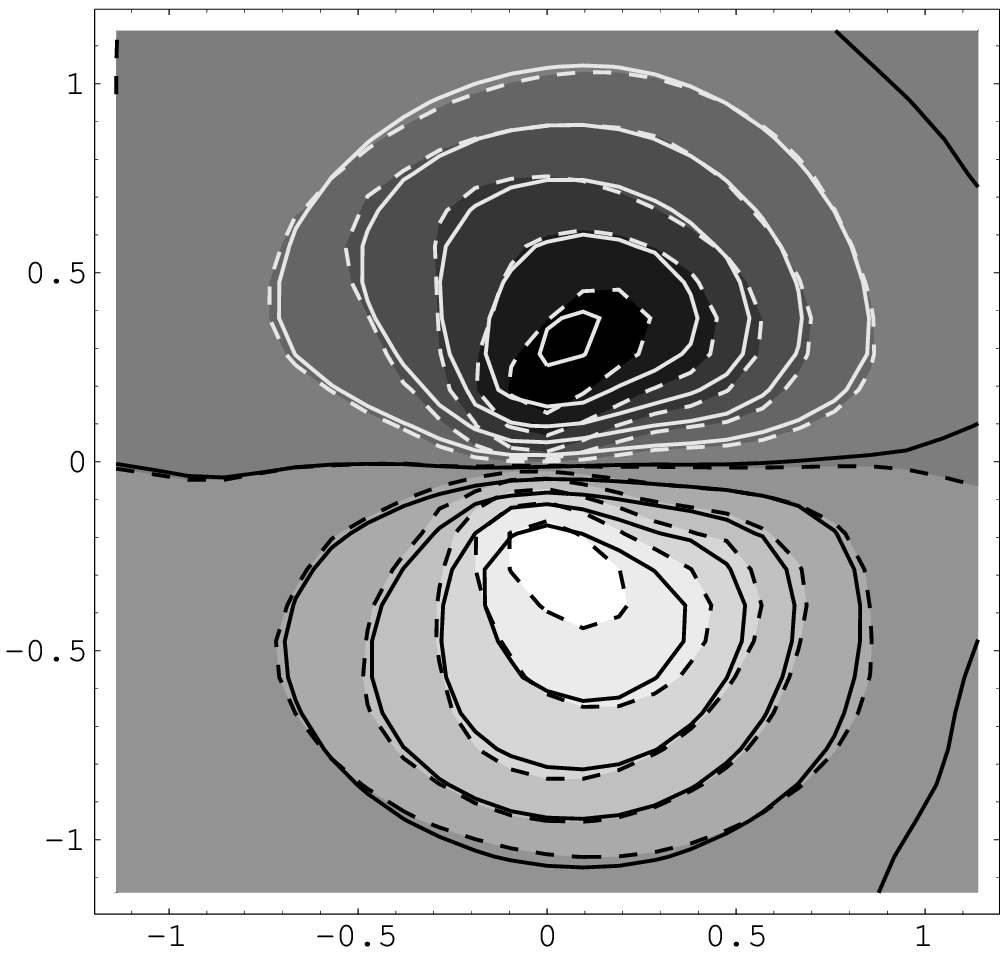,width=0.5\linewidth}
\epsfig{file=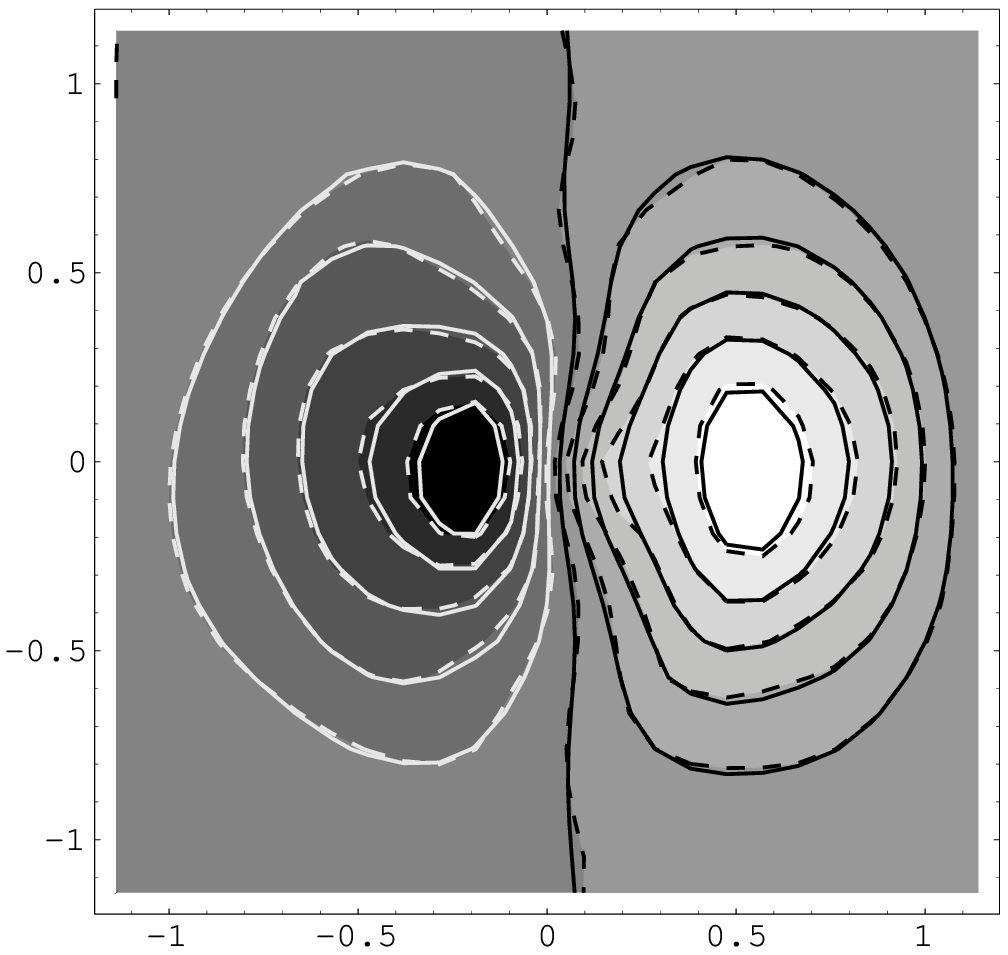,width=0.5\linewidth}}
\centerline{\hspace*{0.2in}(c)\hspace{3.1in}(d)\hfill}
\centerline{\epsfig{file=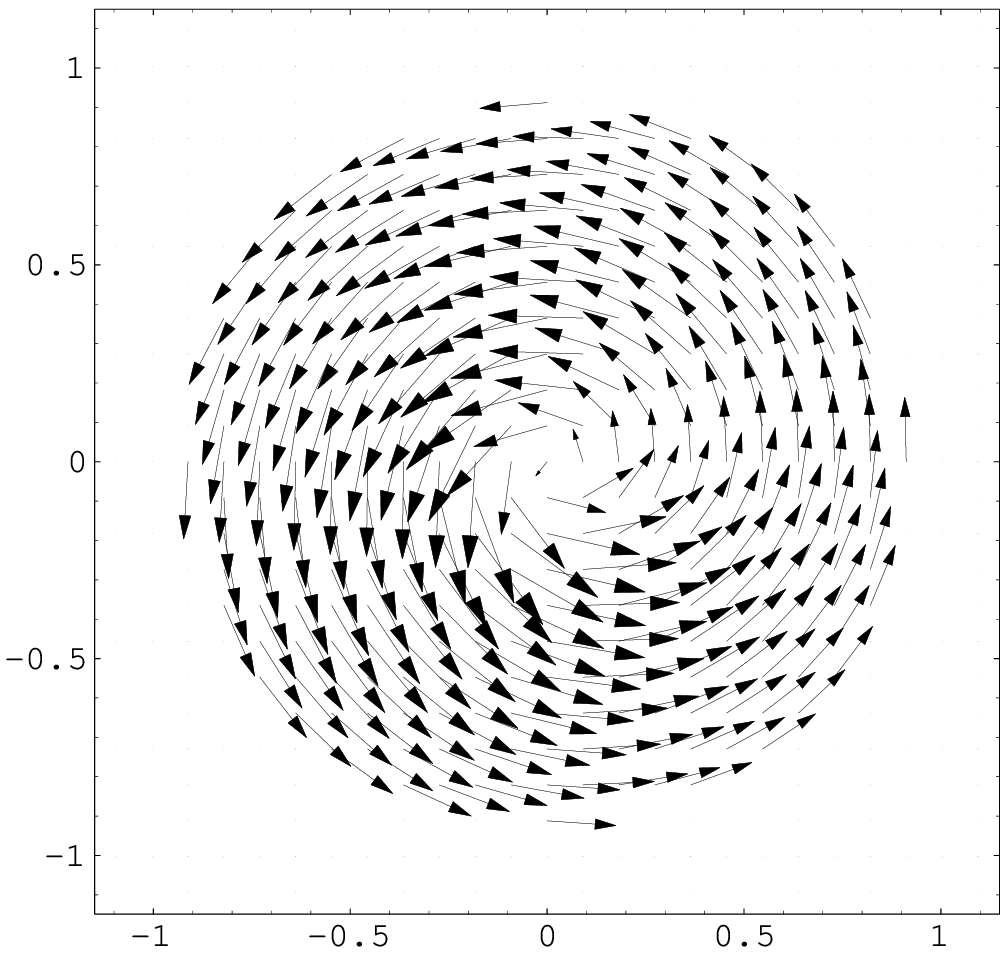,width=0.5\linewidth}
\epsfig{file=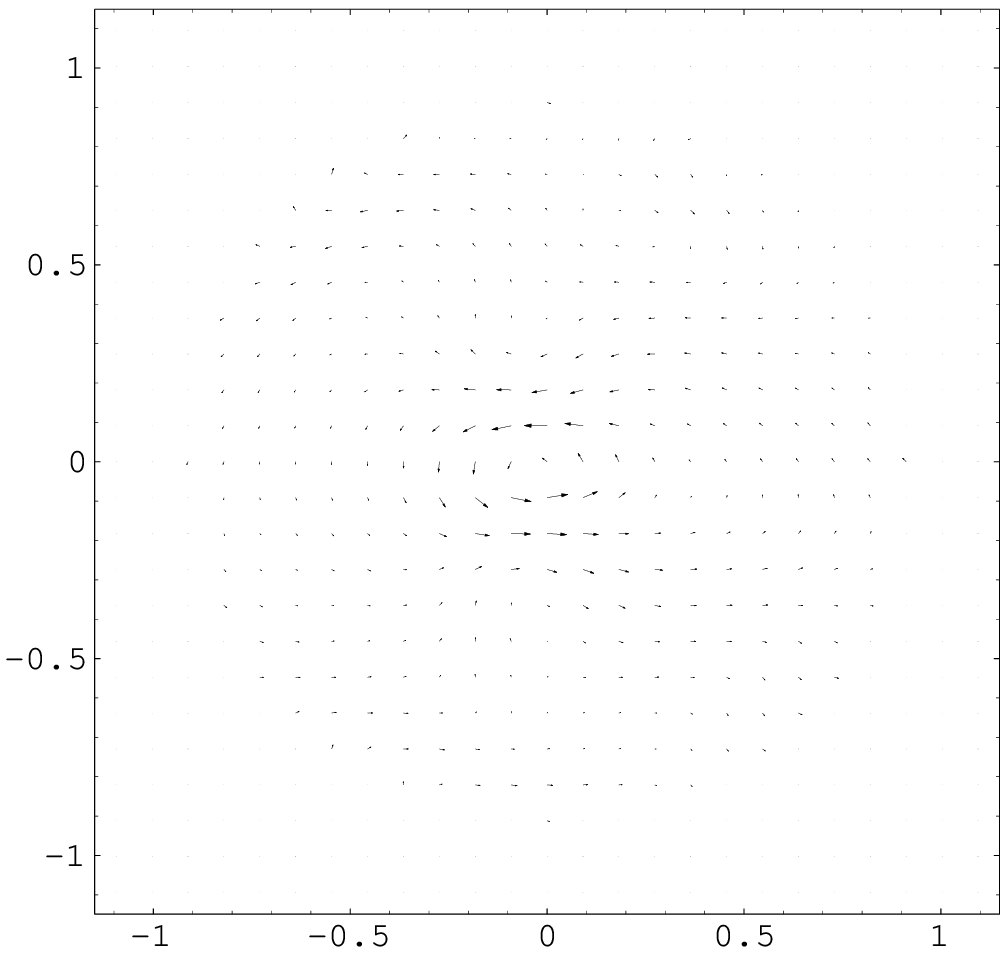,width=0.5\linewidth}}
\caption{Isocontours of the $x$--current densities {\bf (a)}, and
$y$--current densities {\bf (b)} in the disc plane. The contour levels are
equally spaced in current units. The continuous and dashed curves
correspond to the  recovered and model current densities, respectively.
The black and white shadings correspond to negative and positive current
densities. The velocity field of the input model is shown
in {\bf (c)}. The recovered velocity field is close to the input model,
and we show only the residuals (i.e. recovered minus model velocities) in
{\bf (d)}. All axes are in ($''$).}
\label{fig2}
\end{figure}

We computed the integrals in equation~(\ref{twobs}), using  $\Sigma_{\rm
sky}$ and $F_{\rm sky}$ from  Figure~(\ref{fig1}), for eleven different
values of $Y$. Following \citet{gkm99}, we plotted the eleven different
values of one integral against the eleven values of the other integral.
The slope of the ``best--fit'' straight line (in the least--square sense)
gave an estimate of $\omp\sin i$. Using $i = 51\fdg54$, we found that 
$\omp = 15.11\pm 0.47\kmspc\,$. We used this value of $\omp$ to compute
the right side of equation~(\ref{xcurrentobs}). After deprojection using 
$(X, Y) \,=\, (x, y\cos i)\,$ we obtained the $x$--current density,
$\left(\Sigma v_x\right)$, the isocontours of which are displayed in
Figure~(2a) as the continuous curves. For comparison, we also plot the
isocontours of $\left(\Sigma v_x\right)$ from the input model in the same
figure as the dashed curves. In Figure~(2b) similar plots of $\left(\Sigma
v_y\right)$ are displayed to make the point that, in practice deprojection
can also give rise to errors. It is traditional and useful to look
at the velocity field, instead of the current density field. The velocity
field is obtained by dividing the current density field by $\Sigma(x, y)$,
and we may expect this process of division to give rise to errors,
especially in the outer parts where $\Sigma$ is small. To quantify the
errors, we computed the residual--map, which was defined as the difference
between the recovered and input $x$--velocity maps. The $\Sigma$--weighted
mean ($R$) and root--mean--squared (rms; $\sigma_R$) value of the residual
map were then calculated. When expressed in units of
$\left|v_x\right|_{\rm max}$ of the input map, these were found equal to
$R = 1.09\times 10^{-3}$ and $\sigma_R = 7.67\times 10^{-2}$. These
globally determined numbers should give the reader some idea of the
dynamic range of the recovery method, when applied to noise--free spatially
smoothed data. The spatial distribution of the errors in both the $x$ and
$y$ velocities is best visualised with ``arrow plots'' of the velocity
fields. Figure~(2c) displays the velocity field of the input model in the
disc plane. The reconstructed velocities are close to the model, and we do
not present them separately. Instead we plot the residual current field
(recovered minus input) in Figure~(2d).

\begin{figure}
\centerline{\hspace*{0.2in}(a)\hspace{3.1in}(b)\hfill}
\centerline{\epsfig{file=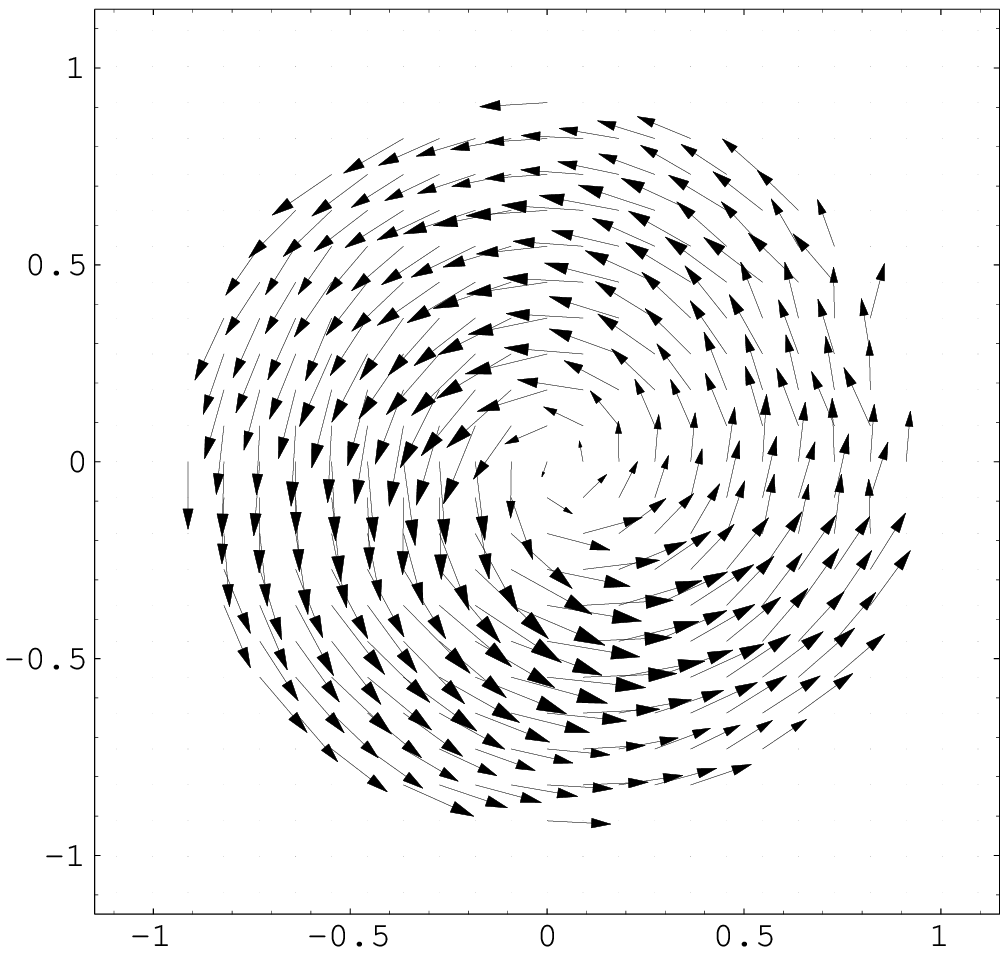,width=0.5\linewidth}
\epsfig{file=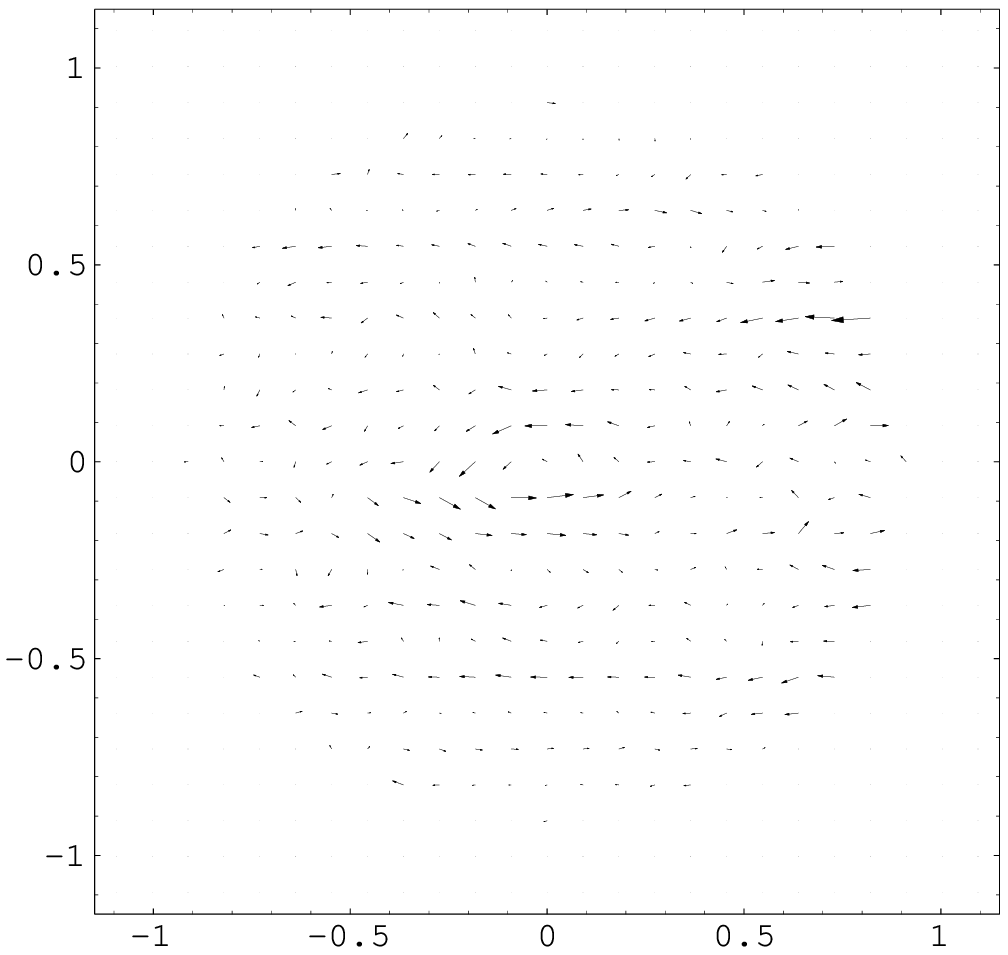,width=0.5\linewidth}}
\centerline{\hspace*{0.2in}(c)\hspace{3.1in}(d)\hfill}
\centerline{\epsfig{file=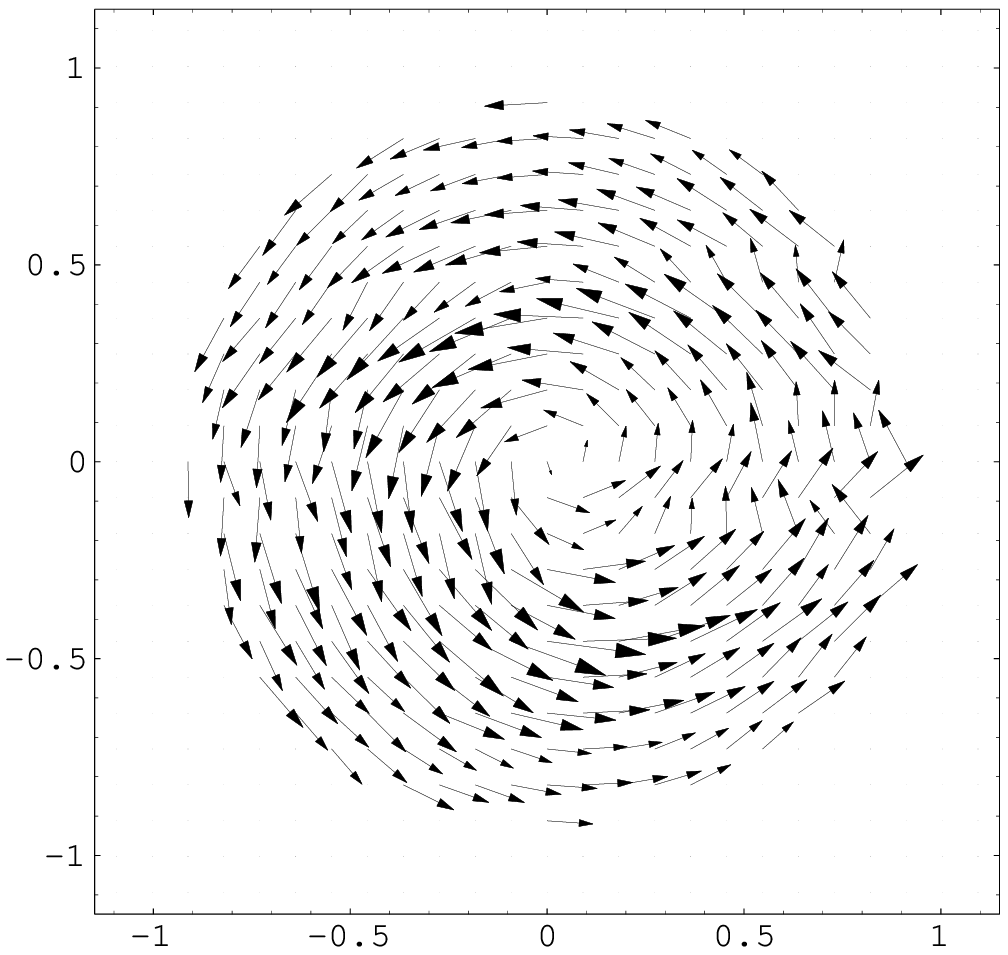,width=0.5\linewidth}
\epsfig{file=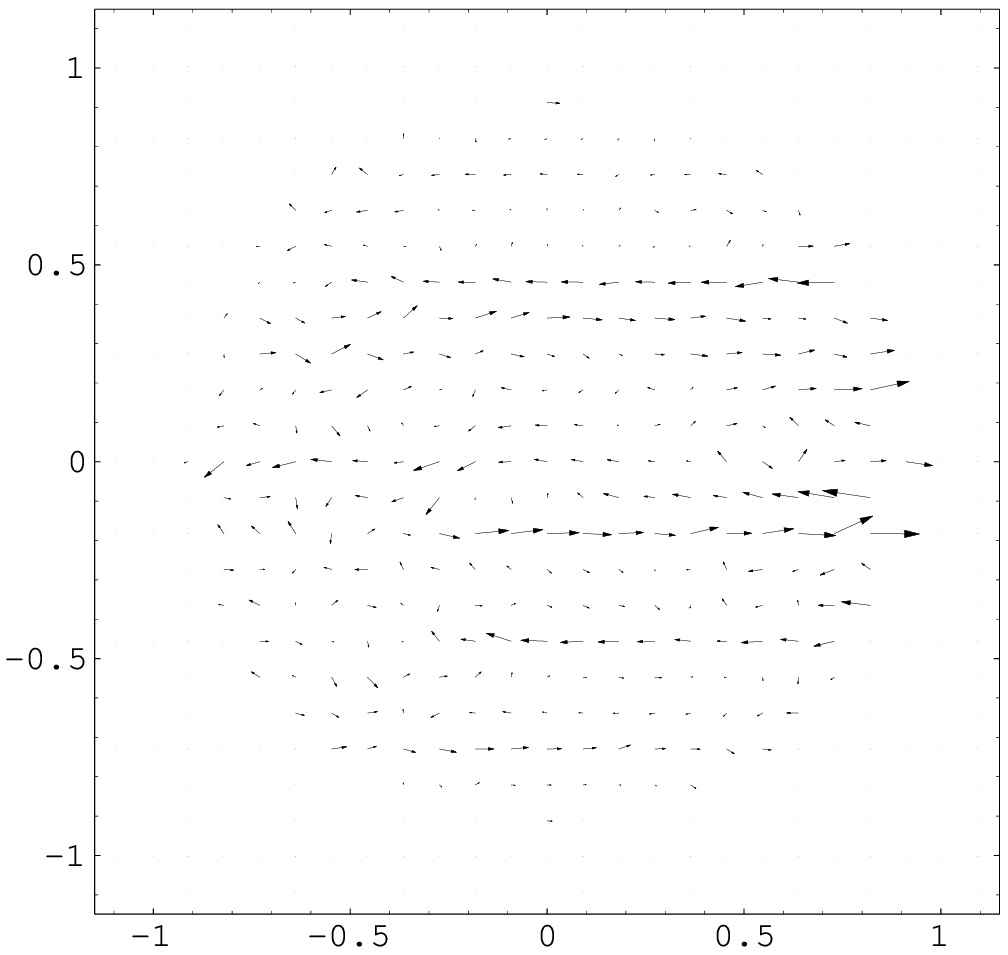,width=0.5\linewidth}}
\caption{Recovered velocity fields and residuals for typical noisy
realisations. {\bf (a)} and {\bf (c)} are the recovered velocity fields
for noise levels of $5\%$ and $10\%$, respectively. {\bf (b)} and {\bf
(d)} are the respective residuals.}
\label{fig3}
\end{figure}

We also tested the recovery method on noisy data. In a real observation
most of the error is likely to reside in the measurement of velocities, 
rather than the surface brightness. This is because the methods used to
extract the velocity information from spectra are less robust than
photometry. Therefore we added noise to Figure~(1b), and kept Figure~(1a)
noise--free. To each pixel of Figure~(1b), we added Gaussian noise with 
mean equal to that observed, and rms equal to some fixed fraction of the
mean.\footnote{Adding noise to $F_{\rm sky}$ is equivalent to adding noise
to $\overline{U}\,=\, (F_{\rm sky}/\Sigma_{\rm sky})\,$, because we have
kept $\Sigma_{\rm sky}\,$ noise-free.}. We experimented with three levels
of noise, namely rms noise per pixel equal to $1\%$, $5\%$, and $10\%$ of
the mean. For each level of noise, twenty one realisations were explored. 
The pattern speed, $x$--current density and $x$--velocities were computed
for each realisation, using  equations~(\ref{twobs}) and
(\ref{xcurrentobs}). Comparing with the input model, we computed the mean
($R$) and rms ($\sigma_{R}$) of the residual $x$--velocities for each
realisation. The distributions of the twenty one $R$'s and $\sigma_{R}$'s
were peaked close to their mean values, $\overline{R}\,$ and
$\overline{\sigma_R}\,$, respectively. The Table below provides estimates
(and rms errors) for these, as well as for the pattern speed.

\begin{table}
\begin{center}
\setlength\tabcolsep{15pt}
\begin{tabular}{|l|c|c|r|} \hline
Noise & $\omp$ (error)& $\overline{R}\;$ (error) & $\overline{\sigma_R}$
(error) \\ \hline
$1\%$ & $15.29\; (0.79)$ & $1.14\times 10^{-3}\; (1.75\times 10^{-4})$ &
$8.02\times 10^{-2}\; (1.54\times 10^{-3})$\\
$5\%$ & $17.67\; (3.24)$ & $1.50\times 10^{-3}\; (6.69\times 10^{-4})$ &
$1.30\times 10^{-1}\; (1.45\times 10^{-2})$\\
$10\%$ & $16.78\; (6.38)$ & $1.00\times 10^{-3}\; (1.22\times 10^{-3})$ &
$2.30\times 10^{-1}\; (3.42\times 10^{-2})$
\\ \hline
\label{table}
\end{tabular}
\caption{Column(1): Noise level added to ``observed'' radial velocity map
The quantities in Columns(2,3,4) were obtained by averaging over
twenty one realisations for each level of noise. Column(2): pattern
speed from using equation~(\ref{twobs}), in units of $\kmspc$.
Columns(3,4): mean and rms residuals of recovered transverse
velocities, respectively, in units of $\left|v_x\right|_{\rm max}$
of the input map.}
\end{center}
\end{table}

\noindent The mean residual, $\overline{R}\,$, is always quite small, 
implying that there is very little global systematic shift in the 
$x$--velocities. This occurs because of cancellation between positive 
and negative residual velocities. The estimated pattern speed is also
well--behaved, because this is calculated using numbers from different $Y$
cuts. However, the errors on $\omp$ increase dramatically with noise,
resulting in a significant increase in $\overline{\sigma_R}\,$. As earlier, 
the arrow plots are very revealing. The recovered and residual maps for
the case of $1\%$ noise are very close to the noise--free case discussed
earlier. Therefore, in Figure~3 we display arrow plots only for noise
levels of $5\%$ and $10\%\,$. In addition to random errors in the residual
velocities there are systematic alignments parallel to the line of nodes,
the axis along which integrals are evaluated in the recovery method.
However, as Figure~3 suggests, even for a noise level as high as $10\%$,
the recovery method does not fail completely.  

\section{CONCLUSIONS}

We have demonstrated that it is possible to recover the transverse
velocities of steadily rotating patterns in flat galaxies, using 
two dimensional maps of a tracer surface brightness and radial current
density, if the tracer satisfies a source--free continuity equation.
Our method is kinematical, and closely related to the TW method of 
determining pattern speeds. Indeed, the conditions that need to be
satisfied---that the galaxy is flat, the pattern is steadily rotating, 
and the tracer obeys a continuity equation---are identical to those 
assumed by TW84. Our main result is an explicit expression for the
transverse velocities (equation~\ref{xcurrentobs}), which is exact under
ideal conditions. We have applied it successfully to simulated data, and
demonstrated its utility in the presence of intrinsic numerical errors in
the data, finite angular resolution, and noise. The TW relation for the
pattern speed (equation~\ref{twobs}) emerges as an eigenvalue, and we
expect our method to work well whenever the TW method gives a good
estimate of the pattern speed. It is legitimate to be concerned that the
conditions required to be satisfied might impose serious limitations in
practice; the angle of inclination and line of nodes need to be estimated,
the pattern need not be steadily rotating, the continuity equation need
not be satisfied, the tracer distribution could be warped, the disc could
be thick, and there could be streaming velocities in the $z$ direction.
All these are well--known worries about the applicability of the TW method
itself. That they are not unduly restrictive is evident from the success
that the TW method itself has enjoyed in the determination of pattern
speeds \citep[see e.g.][]{ken87, kt91, mk95, gkm99, bur99, bak01, dw01,
dca02, dgs02, ger02, zr02, aug03}. Therefore we are cautiously optimistic
that our method of recovering transverse velocities can be applied
usefully to two dimensional spectral maps. 

\section{Acknowledgments}
We are grateful to  D.~Bhattacharya, E.~Emsellem, V.~Radhakrishnan, and 
in particular an anonymous referee for very helpful comments and
suggestions. NS was supported by grant 20-64856.01 of the Swiss National
Science Foundation.

\end{document}